\documentclass[aps,prl,twocolumn,superscriptaddress]{revtex4} 
\usepackage{color}
\usepackage{graphicx}
\usepackage{amsmath,amssymb}
\usepackage{soul}
\usepackage{pdfpages}

\begin{document}

\title{Dependence of DNA persistence length on ionic strength and ion type}

\author{S\'ebastien Guilbaud}
\affiliation{Institut de Pharmacologie et de Biologie Structurale, Universit\'e de Toulouse, CNRS, UPS, France}
\altaffiliation{Present address: Department of Physics, University of York, UK}
\author{Laurence Salom\'e}
\affiliation{Institut de Pharmacologie et de Biologie Structurale, Universit\'e de Toulouse, CNRS, UPS, France} 
\author{Nicolas Destainville}
\affiliation{Laboratoire de Physique Th\'eorique (IRSAMC), Universit\'e de Toulouse, CNRS, UPS, France}
\author{Manoel Manghi}
\email{manghi@irsamc.ups-tlse.fr}
\affiliation{Laboratoire de Physique Th\'eorique (IRSAMC), Universit\'e de Toulouse, CNRS, UPS, France}
\author{Catherine Tardin}
\email{tardin@ipbs.fr}
\affiliation{Institut de Pharmacologie et de Biologie Structurale, Universit\'e de Toulouse, CNRS, UPS, France} 

\date{\today}

\begin{abstract}
Even though the persistence length $L_P$ of double-stranded DNA plays a pivotal role in cell biology and nanotechnologies, its dependence on ionic strength $I$ lacks a consensual description. Using a high-throughput single-molecule technique and statistical physics modeling, we measure $L_P$ in presence of monovalent (Li$^+$, Na$^+$, K$^+$) and divalent (Mg$^{2+}$, Ca$^{2+}$) metallic and alkyl ammonium ions, over a large range 0.5~mM~$\leq I\leq 5$~M. We show that linear Debye-H\"uckel-type theories do not describe even part of these data. By contrast, the Netz-Orland and Trizac-Shen formulas, two approximate theories including non-linear electrostatic effects and the finite DNA radius, fit our data with divalent and monovalent ions, respectively, over the whole $I$ range. Furthermore the metallic ion type does not influence $L_P(I)$, in contrast to alkyl ammonium monovalent ions at high~$I$. 
\end{abstract}

\maketitle

The experimental and theoretical study of polyelectrolyte stiffness has been an active field of research in the last 40 years~\cite{Odijk1977,Skolnick1977,Manning1981,Fixman1982,LeBret1982,Barrat1993,Baumann1997,Wenner2002} because its potential implications in biology, biophysics, and biotechnologies are tremendous. The diverse ionic conditions existing in the intracellular surroundings, in terms of both ionic strength and ion species~\cite{Fagerbakke1996,Heldal1985,Alberts2002},  impact most of the biological macromolecules, particularly the double-stranded DNA (dsDNA), which bears one of the highest negative linear density of charges among biopolymers (2 e$^-$ per base-pair). 
In nanotechnological applications, salt conditions determine the capacity of self-assembling of single stranded DNA as well as the mechanical properties of the resulting nanostructures~\cite{Lee2009,Halverson2013}, e.g., DNA origami~\cite{Goodman2005,Reinhardt2014,Fischer2016,Hong2017} or aptamers~\cite{Padma1993}. Even though various fields of science are concerned, how ionic conditions influence dsDNA stiffness remains controversial from a physical perspective. Stiffness is quantified by the bending persistence length, $L_P$, the tangent-tangent correlation length, which has two contributions: a bare one, $L_P^0= K/(k_{\rm B}T)$, related to the bending modulus $K$~\cite{Kratky}, and an electrostatic one associated with electrostatic repulsion within the polyelectrolyte, which is partially screened as its surrounding is enriched in counterions. As a result, $L_P$ decreases when ion concentration grows; however, strong discrepancies exist between various experimental results obtained with different techniques~\cite{Brunet2015,Savelyev2012}, notably in force-free conditions~\cite{Brunet2015} or in stretching experiments~\cite{Baumann1997,Saleh2009}. Furthermore, as discussed below, theoretical approaches struggle with providing a consensual frame embracing the whole range of ionic conditions.
\begin{figure*}[t!]
\begin{center}
\includegraphics[height=4.4cm]{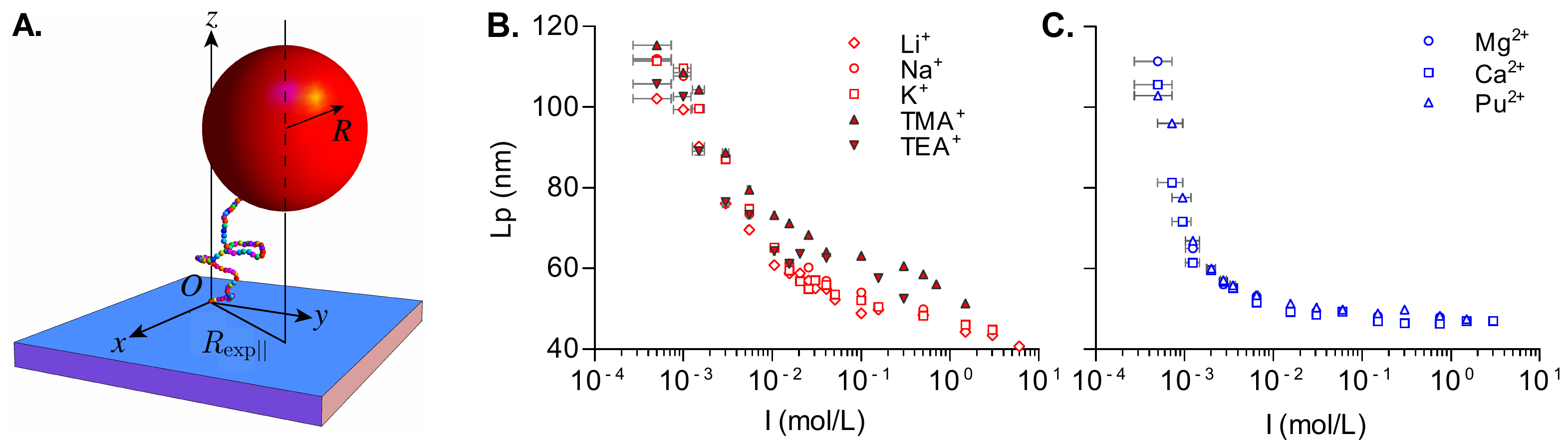}
\caption{(A) Sketch of the Tethered Particle Motion set-up. The measured quantity is $R_{\rm exp||}$. Right: Influence of the ionic strength on the mean persistence length of a 1201 bp DNA for (B) monovalent and (C) divalent cations.
\label{LpvsI:fig}}
\end{center}
\end{figure*}

Accurate experimental quantifications of these $L_P$ changes are indeed non-trivial as the data analysis is not usually straightforward~\cite{Brunet2015}. Recently, we have established a methodology based on high-throughput tethered particle motion (HT-TPM, see Fig.~\ref{LpvsI:fig}A), in which a high density of individual dsDNA molecules are tethered to a micro-patterned surface by one of their extremities, while the other one is labeled with a sub-micrometer-sized nanoparticle (see Supplemental Material SM)~\cite{Plenat2012}. Tracking the nanoparticles thus allows us the monitoring of the conformational dynamics of single dsDNA in almost force-free conditions~\cite{Segall2006}. Following a procedure of analysis based on statistical physics modeling, we established a rigorous method to retrieve $L_P$  from the r.m.s. of the projected end-to-end distance of the tethered particles, $R_{\rm exp\parallel}$~\cite{Manghi2010}, and quantify its decrease as a function of the ionic strength $I=\frac12 \sum_i z_i^2 c_i$  with $z_i$ the valence (in units of the elementary charge $e$) and $c_i$ the concentration of ion $i$ (see Fig.~\ref{LpvsI:fig}).

From a theoretical perspective, the popular Odijk-Skolnick-Fixman (OSF) model~\cite{Odijk1977,Skolnick1977} assimilates the polyelectrolyte to an infinitely thin and rigid rod with a uniform linear density of charges $A^{-1}$ ($=6~e$/nm for dsDNA where $e$ is the elementary charge). The mobile ions, regarded as point-like, organize in space according to the Boltzmann distribution, where the electrostatic potential is determined by linearizing the mean-field Poisson-Boltzmann (PB) equation, in the Debye-H\"uckel approximation valid for low electrostatic potentials. The OSF theory leads to 
\begin{equation}
L_P = L_P^0 + \frac{\ell_{\rm B}}{4 A^2 \kappa^2},
\label{OSF}
\end{equation}
where $\kappa = (8\pi \ell_{\rm B}I)^{1/2}$ is the Debye parameter, $\ell_{\rm B}=e^2/(4 \pi \varepsilon k_{\rm B}T) \simeq 0.7$~nm at 20$^\circ$C in water is the Bjerrum length. Due to the hypothesis of low electrostatic potential, OSF theory is only valid for high $I$, typically above 0.1 M.  At low $I$, a correction to the OSF model was proposed by Manning~\cite{Manning1981}, where part of the ions condense along the DNA so that the distance between the unscreened DNA elementary charges increases up to $z \ell_{\rm B}$. The resulting OSFM model leads to 
\begin{equation}
L_P(I)= L_P^0 +\alpha^2(z) \frac{\ell_{\rm B}}{4 A^2 \kappa^2}
\label{OSFM}
\end{equation}
where the effective fraction of charges along the DNA $\alpha=A/(z \ell_{\rm B})$ depends on $z$. In order to embrace the whole range of $I$ explored experimentally, a model developed by Netz and Orland (NO)~\cite{Netz2003} and adapted in~\cite{Brunet2015}, is based on a variational approximation of the full PB equation. This NO theory leads to a more complicated effective charge $\alpha(z,\kappa R_{\rm DNA})$ that depends on the DNA radius, $R_{\rm DNA}$, and grows with $I$. 
Finally, in 2016, Trizac and Shen (TS) corrected the OSFM formula by taking into account the first term in an expansion in $\kappa R_{\rm DNA}$of the electrostatic potential, and interpolating between exact solutions of the PB equation (in the limits of zero and high salt) for the effective charge of the DNA, $\xi_{\rm eff}$, that also varies with $\kappa R_{\rm DNA}$~\cite{Trizac2016}. Valid only for monovalent ions, it yields the same form as Eq.~\eqref{OSFM} with $\alpha$ replaced by $\alpha= \frac{A\xi_{\rm eff}}{\ell_{\rm B}}(1 + \kappa R_{\rm DNA})^{1/2}$. Hence, the TS formula differs from the NO one by the corrective term and the expression of the effective charge (computed variationally in the NO approach). 

In Ref.~\cite{Brunet2015}, data were obtained following this HT-TPM procedure with Na$^+$ and Mg$^{2+}$ ranging from 10~mM to 3~M and 0.3~M, respectively. The first 3 models were used to fit the data. The OSF and OSFM models could not account quantitatively for the whole experimental data set obtained with Na$^+$ or Mg$^{2+}$. For the range of $I$ studied, a reasonable scaling interpolation of the NO factor was $\alpha \propto (\kappa R_{\rm DNA})^{\beta(z)}$ where $\beta(z)$ is an effective exponent. The NO approach could then fit the Mg$^{2+}$ data only, while the Manning stretching model~\cite{Manning2006}, which incorporates the internal stretching of the polymer modified by ion screening, succeeded in fitting the Na$^+$ data only. 

In this Letter, challenging further the existing theories predicting $L_P(I)$, we examine a 1201~bp dsDNA (i)~on an extended range of $I$ down to 0.5~mM and up to 6~M  (under well-controlled $p$H comprised between 7 and 7.3) and (ii)~with a variety of ions with different ion-specific characteristics (such as radius or hydrophobicity), neglected in all the existing theories~\cite{Manning2015} (see Fig.~\ref{LpvsI:fig}). We took much care to evaluate the influence of a large set of biologically and biotechnologically relevant ions: Li$^+$, Na$^+$, K$^+$, tetramethyl ammonium TMA$^+$, tetraethyl ammonium TEA$^+$, Mg$^{2+}$, Ca$^{2+}$,  putrescine Put$^{2+}$ (see SM Table~1). We confirm that neither the OSF theory nor its Manning refinement (OSFM) describe even part of the data. By contrast, the NO model and the TS one are shown to fit accurately the data obtained with the chosen divalent and monovalent ions respectively and up to $I=1$~M, with reasonable values for the fitting parameters $L_P^0$ and $R_{\rm DNA}$. We therefore demonstrate in this work that the radii of metallic ions do not influence $L_P$ except in the case of large alkyl ammonium monovalent ions, for which a distinct $L_P^0$ at high salt is obtained.
\begin{figure*}[t!]
\begin{center}
\includegraphics*[height=8cm]{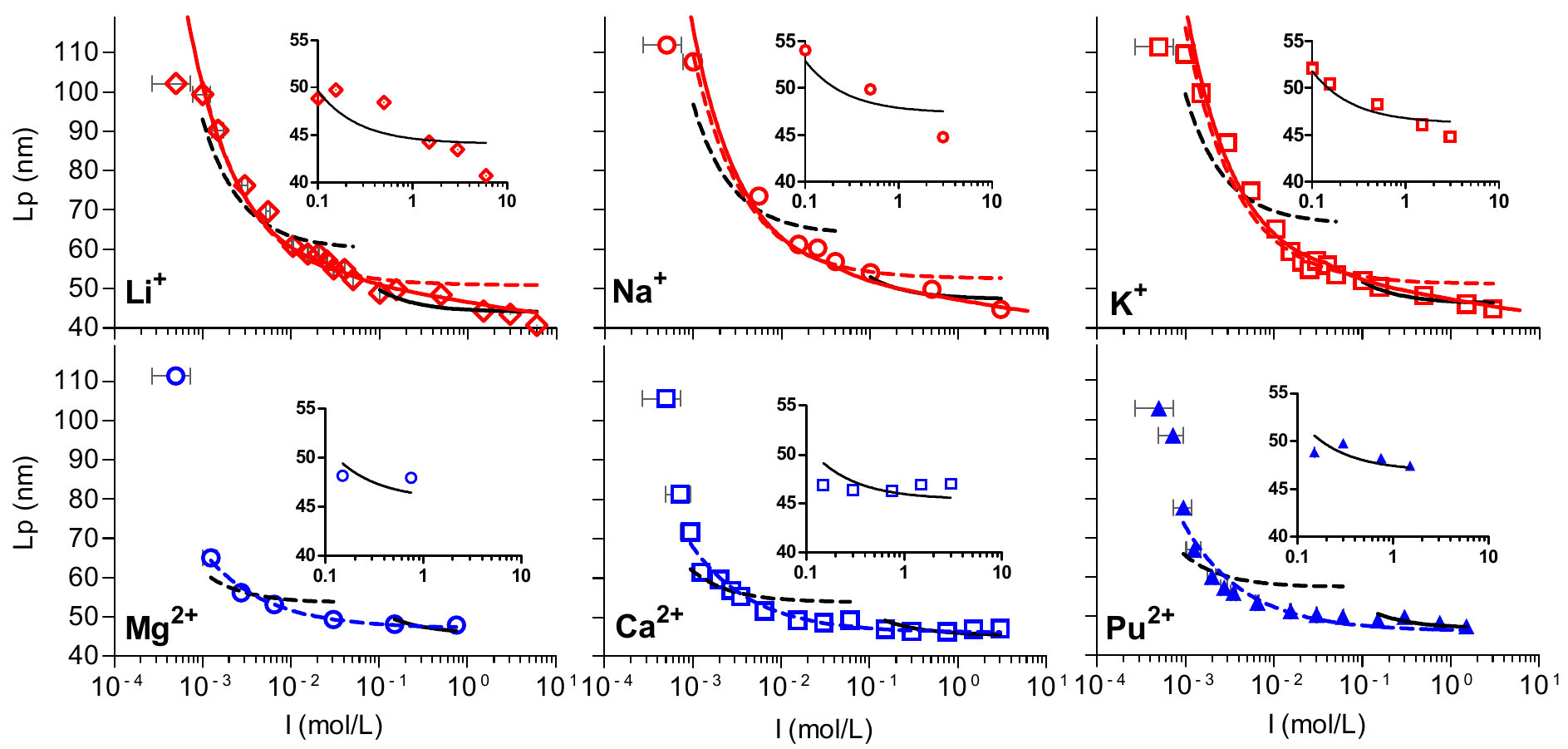} 
\caption{Ionic strength dependence of $L_P$ for monovalent (top, red) and divalent (bottom, blue) ions with fitting curves corresponding to OSF ($L_P(I)= L_P^0 + 0.559/I$, where $I$ is expressed in mol/L and $L_P$ in nm; black line), OSFM (black dashed line), NO (dashed lines) and TS (solid lines) theories. Insets are zooms on the high $I$ region. For monovalent ions, the NO model leads to $L_P = L_P^0 + C R^{0.490}_{\rm DNA} I^{-0.755}$ considering that $\alpha = 0.635 (\kappa R_{\rm DNA})^{0.245}$  when $0.2 \leq \kappa R_{\rm DNA})^{0.245} \leq 1.5$, i.e. $ 4 \leq I \leq 208$~mM for $R_{\rm DNA}=1$~nm. For divalent ions, it leads to $L_P = L_P^0 + C' R^{0.728}_{\rm DNA} I^{-0.636}$ considering that $\alpha = 0.423 (\kappa R_{\rm DNA})^{0.364}$ when $0.2 \leq \kappa R_{\rm DNA})^{0.245} \leq 2.5$, i.e.  $ 4 \leq I \leq 577$~mM.
\label{fits}}
\end{center}
\end{figure*}

To explore the influence of ions with this extended range of $I$ on $R_{\rm exp ||}$, we thoroughly considered the buffer composition and the influence of $p$H, which is often neglected.
We became aware that, even at the usual concentration of phosphate buffer, $p$H decreased when ions were added (see SM). This occurred moderately for monovalent ions but quite dramatically for divalent ions (for $I>0.5$~M). Consequently, instead of using a phosphate buffer, we chose an HEPES buffer at 1 mM $p$H 7.4, denoted zero-salt buffer (with a minimum ionic strength of 0.5~mM).
Using it $p$H is maintained between 7 and 7.3. The results obtained in presence of Na$^+$ do not exhibit any fall at high $I$, as seen in Ref.~\cite{Brunet2015}; similarly, those obtained in presence of Mg$^{2+}$ show a much less pronounced slope. This slower decrease in $R_{\rm exp ||}$ at high $I$ is clearly correlated with the improved $p$H stabilization obtained in 1~mM HEPES buffer, as we experimentally confirmed that acidic $p$H negatively affected $R_{\rm exp||}$.

We then supplemented the zero-salt-buffer with monovalent metallic ions Li$^+$, Na$^+$, K$^+$ (ionic radii ranging from 0.071 to 0.141~nm), and divalent metallic ions Mg$^{2+}$ and Ca$^{2+}$ (ionic radii of 0.070 and 0.103 nm, see SM Table 1). When $I$ increases from 0.5 mM to 3 M, $R_{\rm exp \parallel}$ decreases by about 20\% for both type of metallic ions (SM Fig.~2). We notice a faster decrease for divalent ions. In order to consider much larger ions, we carried out experiments with three polyamines TMA$^+$, TEA$^+$, and Put$^{2+}$ (an essential metabolite of many living organisms, e.g., found at high concentration in {\em E. coli}~\cite{Miller2015}). TMA$^+$ and TEA$^+$ have radii 3 to 4 times larger than those of the chosen metallic monovalent ions. Put$^{2+}$ size has not been characterized yet; however, its radius likely exceeds that of the metallic ions. Surprisingly, in the case of divalent ions, $R_{\rm exp \parallel}$ values are extremely similar for both metallic ions and Put$^{2+}$. In presence of TMA$^+$ and TEA$^+$, they are significantly higher than those obtained with metallic monovalent ions. 

From $R_{\rm exp \parallel}$, we extracted $L_P$ using calibration curves obtained by exact sampling simulation based on a statistical physics model of DNA~\cite{Brunet2015}. As expected, $L_P$ decreases much faster when $I$  increases in presence of divalent ions than in presence of monovalent ions. Unexpectedly, the data superimpose on a unique curve for the three divalent ions (Fig.~\ref{LpvsI:fig}C) but not for the five monovalent ones (Fig.~\ref{LpvsI:fig}B). In addition, $L_P$ reaches a plateau above $I \approx 50$~mM for the divalent ions while, in the case of monovalent ions, only a shoulder at $I \approx 200$~mM can be detected within the continuously decreasing curve. Molecular dynamics simulations examining the role of Na$^+$ identified a similar transitory plateau followed by a fall at high salt (SM Fig.~5)~\cite{Savelyev2012,Savelyev2010}.
Moreover, we have measured that in the presence of monovalent ions at  $I \approx 150$~mM, adding even only 1~mM divalent ions leads to a significant decrease of  $\approx 6$~nm in $L_P$  (SM Fig.~4). This demonstrates an additive effect of monovalent and divalent ions at these biologically relevant ion concentrations.

To determine which theory best describes our experimental results, we performed fits with four equations corresponding to the OSF, OSFM, NO and TS models. The fitting curves are displayed in detail in Fig.~\ref{fits} and in SM (fit parameters are given in SM Tables 2 and 3). For $I\geq100$~mM, we fitted the data with the OSF formula that predicts a saturation at high salt. We observed discrepancies for monovalent ions, which was expected in absence of saturation in the experimental data, as well as for divalent ions, which was less expected since the saturation is observed experimentally. For $I\leq100$~mM, fits of $L_P$ using the OSFM equation was equivalently inadequate for monovalent and divalent ions. We then employed the NO model on the entire $I$  range, excluding the very first points at low $I$ that may be partially biased due to possible plastic tube contamination (see SM) but could strongly contribute to the fit due to their high $L_P$ value, exceeding 100~nm. For the divalent ions, we observed extremely good adjustments of $L_P(I)$ by the NO fits. For monovalent ions, the NO fits also seem visually reasonable for  $1 \leq I \leq 30$~mM; however, the fitting values for $R_{\rm DNA}$ are only half the expected size of 1~nm.
To circumvent this discrepancy as well as the poor fitting at high salt, we considered the TS analytical formula~\cite{Trizac2016}, which considerably improved the $L_P(I)$ fit as observed in Fig.~\ref{fits}. Note that the fitted value $L_P^0=41$~nm is the same for the 3 metallic ions and $R_{\rm DNA}$ is almost constant, between 0.85 and 1~nm. The strong agreement at high $I$ comes from the precise expression of $\xi_{\rm eff}$ at large ionic strength~\cite{Aubouy2003}. The variational theory, on the other hand, looks for the optimized formula $\alpha(\kappa R_{\rm DNA})$ for the whole $I$ range, at the expense of this high precision for large $I$. It is clearly sufficient for divalent ions (for which no TS formula exist) but not for monovalent ones. In particular the NO approach does not perfectly fit the monovalent salt data because its limiting behavior at high $I$ is, by construction, the OSF formula. For the TMA$^+$ and TEA$^+$ ions the value of $L^P_0$ is slightly higher (51 and 47 nm, respectively), which could be correlated to their large size (see SM Table 1).

\begin{figure}[t!]
\begin{center}
\includegraphics*[width=7.5cm]{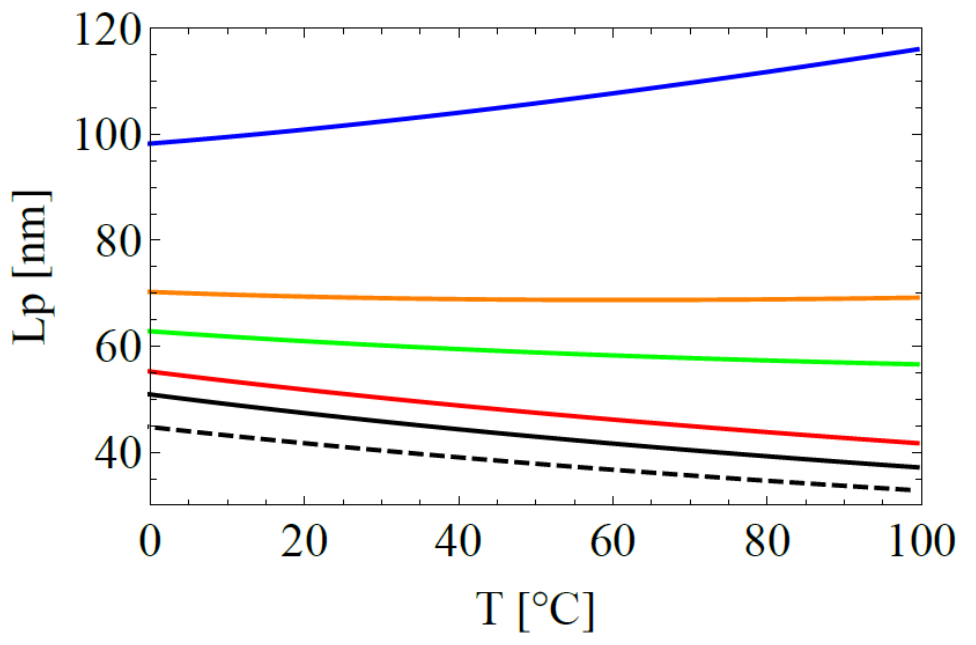} 
\caption{Theoretical prediction of the influence of the temperature $T$ on $L_P$ for monovalent ions and $I=0.001;0.004;0.01;0.1;1$~M from top to bottom. The dashed curve is the infinite $I$ limit showing the non electrostatic contribution $L_P^0=K/(k_{\rm B} T)$.
\label{LpvsT:fig}}
\end{center}
\end{figure}

On the basis of the TS theory, we can also explore the combined effects of the temperature $T$ and the ionic strength $I$ on $L_P$ on structurally intact dsDNA. In a previous work, the effect of the temperature has been measured experimentally at fixed physiological salt conditions $I=160$~mM~\cite{Brunet2018}. It has been shown that $L_P$ decreases as  $1/T$ as expected from the simple formula valid for neutral worm-like chains $L_P^0 = K/(k_{\rm B}T)$. However, not only the bare persistence length $L_P^0$ but also the electrostatic contribution depends on $T$, since entropic effects control the ionic screening of the dsDNA. Hence using our fitting values obtained at $T=20^\circ$C, we plotted $L_P(T)$  for various $I$ in Fig.~\ref{LpvsT:fig}. We observed that for $I \geq 100$~mM the electrostatic contribution is small as compared to $L_P^0$. Therefore, we observe a $1/T$ law with a shift of the curve to lower values when $I$ increases. For $I < 4$~mM, however, we predict a striking reversal with $L_P$ increasing with $T$. New experiments exploring the dependence of the $L_P$ as a function of $T$ and $I$ are therefore needed to check further the theoretical TS approach. 
 
No influence of the size or the nature of the ions was seen as the curves obtained with the three divalent ions and the three metallic monovalent ions superimposed in two unique curves. This complete superimposition prevents us from considering as significant the slight changes of dsDNA radius derived from the fits. The independence of the ionic-strength variation of $L_P$ with the ion size, and our fitted values for $R_{\rm DNA} \approx 1$~nm  are in good agreement with Gebala and coworkers' results~\cite{Gebala2016}. They showed that the atmosphere occupancy around dsDNA by monovalent ions did not depend on the ion size across the monovalent metallic ions except for Li$^+$\cite{Gebala2016}, for which we only observe a slight reduction as compared to the other monovalent metallic ions. These two distinct approaches thus support the same view of an identical behavior for various metallic monovalent ions with different sizes. 
Therefore, in timescales of seconds, the difference in the ion binding sites along the DNA tube~\cite{Varnai2004} and residence times~\cite{Pan2014} measured between Na$^+$ and K$^+$  using  molecular dynamic simulations of tens of nanoseconds, fades away. 

Concerning alkyl ammonium ions, their significantly higher $L_P^0$ suggests that their 3 to 4 times wider ion size precludes a sufficiently large density of ions in the close proximity of dsDNA to completely screen the electrostatic interactions even at large $I$. The capacity of these ions to easily dehydrate due to their disorganized hydration shell and consequently enter the dsDNA groves, as predicted by molecular dynamic simulations~\cite{Bhowmik2014}, is insufficient to ensure an efficient screening of the dsDNA charges. Hydrated divalent metallic ions such as Mg$^{2+}$ were predicted to exhibit a much more localized distribution than Na$^+$ and K$^+$ and spend long resident times of a few nanoseconds within the dsDNA tube~\cite{Pan2014}; yet, we did not measure any impact on the fitted $R_{\rm DNA}$ value. Surprisingly, Put$^{2+}$ behaves as metallic divalent ions in spite of its long linear structure. This is at odds with what was found to describe thermal DNA denaturation in presence of Put$^{2+}$~\cite{Miller2015}~\footnote{Our measurements are performed on 1201~bp long DNA molecules for which any sequence effect~\cite{Varnai2004,Ponomarev2004} is assumed to be self-averaged.}.
 
The great efficiency of NO and TS theories comes from the consideration of non-linear electrostatic terms and of the finite radius of dsDNA. It suggests that they could be also valid for other polyelectrolytes. For hyaluronic acid (HA) in the presence of Na$^+$ in stretching experiments with magnetic tweezers~\cite{Berezney2017}, $L_P$ decreases following $L_P = L_P^0 + {\rm Const.} I^{-\delta}$  with $\delta=0.65$ at low salt to be compared with our value of 0.75, while HA is much more flexible than DNA, with $L_P^0$ about 10 times shorter than the dsDNA one, and bears a reduced charge density of $1~e$/nm for HA vs $6~e$/nm for dsDNA. The use of NO and TS theories to finely model the flexibility of biopolymers such as single stranded RNA~\cite{Chen2012} or chromatin fibres should be extremely useful for the elucidation of gene expression and 3D organization of chromosomes~\cite{Zhu2018} and for the control of the shape of nucleic acid nanostructures~\cite{Liu2018}.\\

\begin{acknowledgments}
We acknowledge Philippe Rousseau (LMGM, CBI, Toulouse) for the DNA samples.\\
\end{acknowledgments}

\newpage
\phantom{a}
\newpage


\section*{Supplementary material}

# Supplemental Material: Dependence of DNA persistence length on ionic strength and ion type

Sébastien Guilbaud[1†], Laurence Salomé[1], Nicolas Destainville[2], Manoel Manghi[2*], Catherine Tardin[1*]

[1] *Institut de Pharmacologie et de Biologie Structurale, Université de Toulouse, CNRS, UPS, France*
[2] *Laboratoire de Physique Théorique (IRSAMC), Université de Toulouse, CNRS, UPS, France*

(Dated: October 2, 2018)

## MATERIAL AND METHODS

### HT-TPM experimental procedure

The DNA sample was produced by polymerase chain reaction amplification (oligonucleotides from Sigma-Aldrich) using oligos Biot-F1201 5-CTGGTGAGTACTCAACCAAG-3 and Dig-R1201 5- CTACAATCCATGCCAACC-3' on pTOC1 plasmid. We use a similar HT-TPM procedure as published in [1]. In brief, coverslips are epoxydized, then micro-contact printed with neutravidin (Invitrogen) to form a square array of isolated spots of about 800 nm size separated by about 3 $\mu$m. The coverslip is assembled with a drilled PEGylated glass slide and, a silicone spacer that forms fluidic channels of about 15 $\mu$L. The internal surface of the chamber is passivated during 10 min with the zero-salt-buffer composed of 1 mM HEPES set at $p$H 7.3 by addition of NaOH, pluronic F127 1 mg/mL and BSA 0.1 mg/mL (Sigma-Aldrich). A 1:1 mix of 50 pM DNA and 300 nm-sized polystyrene particles (Merck) coated with anti- digoxigenin (Roche) is incubated during 30 min at 37 C and is injected in the channels for an overnight incubation.

The channel under study was extensively rinsed ($\sim$ 100 chamber volumes) with the zero-salt-buffer, then with X-salt-buffer composed of the zero-salt-buffer supplemented with various concentrations of X ions (SM Tables IV and V). The salts employed are NaCl (Normapur VWR), KCl (60128 Sigma), tetramethylammonium chloride denoted by TMA$^+$ (87718 Sigma), tetraethylammonium chloride denoted by TEA$^+$ (T2265 Sigma), MgCl$_2$ (M1028 Sigma), CaCl$_2$ (C5080 Sigma-Aldrich), putrescine dihydrochloride denoted by Put$^{2+}$ (P7505 Sigma). More precisely, after addition of X-salt-buffer, the channel was left to incubate for 4 min, the acquisition was performed and the channel was again extensively rinsed ($\sim$ 100 chamber volumes) with the zero-salt-buffer before any new salt condition was applied. Channels under examination underwent first an increase in concentration of monovalent ions then of divalent ion concentration or only one of the two types of salt. We ensured the reliability of the experimental procedure by checking the agreement between the two measurements in the zero-salt buffer obtained before the addition of monovalent ions and before the addition of divalent ions. Experiments were repeated on different days to ensure the reproducibility of our results. A few experiments with TMA$^+$ were performed using PEGylated coverslips. This was done by an overnight silanization of the glass with (3-Mercaptopropyl)trimethoxysilane (Sigma-Aldrich) followed by the injection of a 10:1 mix of PEG-maleimide and Biotin-PEG-maleimide (Sigma-Aldrich) in HEPES 10 mM pH 7. These surfaces were patterned by microcontact printing as described above. All the employed buffers were similar to the zero-salt-buffer or X-salt-buffer except for their deprivation in pluronic. We measured the overnight change in conductivity of a water solution when stored in a plastic tube to be equal to 10 $\mu$S/cm which corresponds to a change of about 0.28 mM considering that the ionic strength depends linearly with the conductivity for this very dilute solution [2]. We selected this value as the uncertainty on the ionic strength.

The acquisitions are performed on a dark-field microscope, Zeiss Observer200 with a x32 objective, equipped with a CMOS camera Hitachi FM200WCL (pixel size = 5.5 $\mu$m) and a temperature-controlled stage set at 25°C, Thermo Plate (Tokai Hit). Videos of 1 min were acquired at 100 Hz with an exposure time $T_{\mathrm{ex}} = 10$ ms. The centroid calculation-based particle tracking gives access to the 2D particle trajectories. Then, are calculated the anchoring points of the DNA-tethered particles, the instantaneous projected 2D distances of these DNA-tethered particles to their anchoring point, the symmetry factors as in [3] and finally the root-mean-square of this distance, denoted by $R_{\mathrm{exp}||\mathrm{raw}}$, determined over a sliding window of 2 s.

### Data analysis

$R_{\mathrm{exp}||\mathrm{raw}}$ is processed according to the procedure described in detail in [4]. In summary, DNA-particle complexes that do not respect the criteria of validity, due in particular to aberrant $R_{\mathrm{exp}||\mathrm{raw}}$ and out of range symmetry factor (for

example induced by a stuck particle or a particle tethered by more than 1 DNA molecule), are discarded. Then the relaxation time $\tau_{||}$ is extracted from the correlation function in $X$ and $Y$ positions as described in [8]. Correction from the detector-averaging blurring effect is performed considering $T_{\text{ex}}$ and $\tau_{||}$ to obtain $R_{\text{exp}||}$. At last, the experimental value of $R_{\text{exp}||}$ of an ensemble of particles is calculated as the mean value of the distribution and the error on $R_{\text{exp}||}$ of an ensemble of particles are obtained by using the bootstrap method of R software (R Foundation for Statistical Computing, Vienna, Austria). The persistence length, $L_P$, was extracted from $R_{\text{exp}||}$ using the calibration curve obtained by exact sampling simulation as published in [7].

**Ions studied in this work**

| Ion | Radius (nm) |
|---|---|
| Li$^+$ | 0.071 |
| Na$^+$ | 0.097 |
| K$^+$ | 0.141 |
| TMA$^+$ [(CH$_3$)$_4$N$^+$] 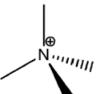 | 0.347 |
| TEA$^+$ [(C$_2$H$_5$)$_4$N$^+$] 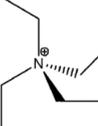 | 0.400 |
| Mg$^{2+}$ | 0.070 |
| Ca$^{2+}$ | 0.103 |
| Put$^{2+}$ [NH$_3^+$(CH2)$_4$NH$_3^+$] 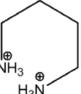 | - |

TABLE I: Radii of the chosen ions [5, 6].

**BUFFER AND $p$H CONDITIONS**

In Ref. [7], the use of a phosphate buffer composed of 1 mM KH$_2$PO$_4$ and 3 mM Na$_2$HPO$_4$ had led to a minimal ionic strength $I = 10.5$ mM when no salt is added. Although biologically relevant, the phosphate buffer, which operates inside cells, has a weak buffering capacity. We could not decrease its concentration and keep a stable neutral $p$H (see SM Fig. 1A). Indeed, we became aware that, even at the initial concentration of phosphate buffer, $p$H decreased when ions were added. This occurred slightly in the case of monovalent ions but quite dramatically in the case of divalent ions at concentrations above 0.5 M. We therefore changed the buffer to an HEPES buffer at 1 mM $p$H 7.4 with buffering capacities that we verified to be maintained even at high concentrations of divalent ions (SM Fig. 1A). We verified that, in this new buffer condition, the amplitudes of motion $R_{\text{exp}||}$ measured in presence of Na$^+$ and Mg$^{2+}$ at low $I$ were in good agreement with the previously published data of Ref. [7], though a few nanometers higher (SM Fig. 2 Left). This slight increase in $R_{\text{exp}||}$ presumably arises from an improvement in the correction of the detector-time-averaging effect on $R_{\text{exp}||}$ as, in these series of experiments, the acquisition time was reduced from 40 to 10 ms [8].

We measured that decreasing the $p$H of a 1 mM HEPES buffer from 7 to 6 results in the decrease of $R_{\text{exp}||}$ by 10 nm. Below $p$H 6, $R_{\text{exp}||}$ collapses, which is followed by massive apparent immobilization of the DNA-particles complexes at $p$H 5.5 (SM Fig. 1B). This collective immobilization was found to be reversible with a return to neutral conditions (SM Fig. 1B). As expected, the strong decrease in $R_{\text{exp}||}$ observed in Ref. [7] between $10 \leq I \leq 15$ mM with Mg$^{2+}$ is now absent. This stems from the binding competition of Mg$^{2+}$ against monovalent ions of the phosphate buffer. A similar bias might appear here for $0.5 \leq I \leq 0.75$ mM, we thus excluded these points from any fit.

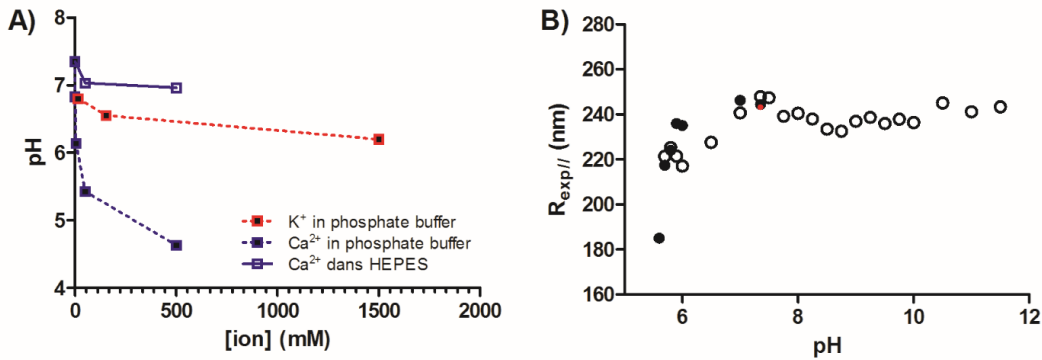

FIG. 1: Causes and consequences of $p$H changes. (A) $p$H decreases with the ion concentration in phosphate buffer (1 mM $KH_2PO_4$ and 3 mM $Na_2HPO_4$) and HEPES (1 mM). (B) Change in $R_{\exp\|}$ as a function of $p$H in solutions composed of 1 mM HEPES buffer and NaOH to adjust $p$H on epoxydized coverslips (filled circles), and on PEGylated coverslips (empty circles). The red filled circle represents the $R_{\exp\|}$ value measured at $p$H 7.35 on a DNA sample that had been previously submitted to $p$H 5.5. It shows the reversibility of the apparent immobilization caused by low $p$H.

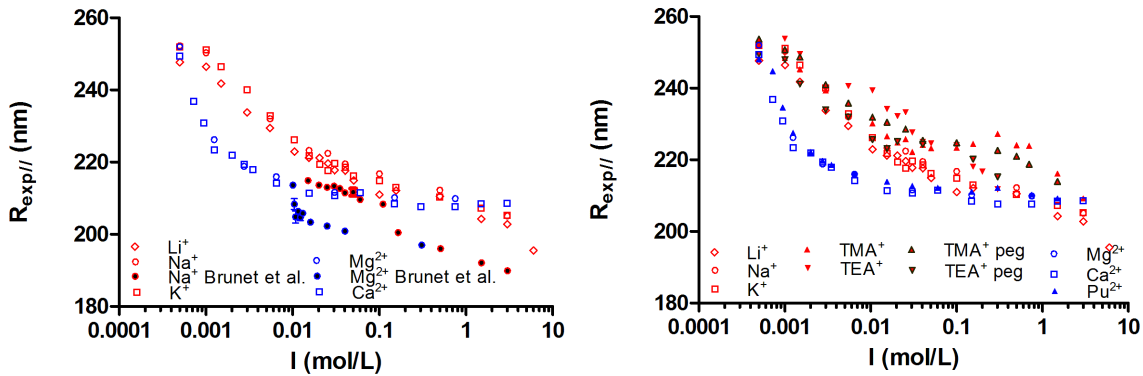

FIG. 2: Influence of the concentrations of ions on the amplitude of motion of 300 nm-sized particle tethered by 1201 bp DNA. Left: metallic ions only and comparison with Ref. [7]. Right: all ions.

## PASSIVATION OF THE BIOCHIP SURFACE

In the case of divalent ions, $R_{\exp\|}$ values are extremely similar for both metallic ions and $Put^{2+}$. In presence of $TMA^+$ and $TEA^+$, they are significantly higher than those obtained with metallic monovalent ions. We wondered whether it could stem from a loss of free ions consecutive to their partial encapsulations into micelles of copolymers used to passivate the biochip surface. To test this possible bias, we repeated the experiments on PEGylated surfaces in absence of any copolymer. We observed that $R_{\exp\|}$ was very little changed in the case of $TMA^+$ but was notably reduced for $TEA^+$ especially at low ionic strength, which could be explained by the presence of additional $TEA^+$ that are not trapped by the copolymer (SM Fig. 3). In this work, we only consider the results obtained with $TMA^+$ and $TEA^+$ on PEGylated biochip surfaces.

Optimizing conditions for this extended study on the influence of $I$ on $L_P$, we became aware of the sensitivity of our measurements with $p$H and the resulting necessity to control it tightly (see above). We were particularly challenged by the fall of $R_{\exp\|}$ when the $p$H was observed to drop below 6.0 in 1 mM HEPES buffer. This fall could be due to technical biases or reveal DNA modifications. The passivating coating of the surface was not involved as adsorbed PEG copolymer or grafted PEG gave the same fall but we cannot exclude the onset of unspecific binding of dsDNA to the antibody-coated particles with acidic $p$H. The dsDNA itself could undergo structural changes that could increase the apparent flexibility of dsDNA. We could not detect any $p$H-driven dsDNA denaturation in UV spectrometry. Our observation may be related to the protonation at acidic $p$H of the cytosines. These changes in the protonation state are known to allow the formation of i-motif or triplex [10], it could also induce a reduction in the internal electrostatic

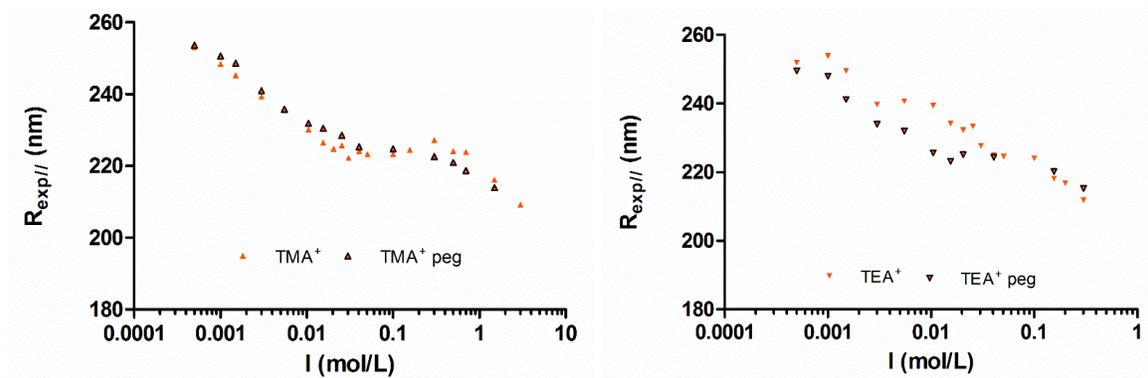

FIG. 3: Influence of the concentration of TMA$^+$ (Left) and TEA$^+$ (Right) on the amplitude of motion of 300 nm-sized particle tethered by 1201 bp DNA in presence of a passivating layer made of adsorbed copolymer (orange triangles) or a grafted PEG layer (orange triangles with black outline).

repulsion of our 50%-GC-rich DNA sample and contribute to this dramatic fall.

## ADDITIONAL FIGURES AND TABLES

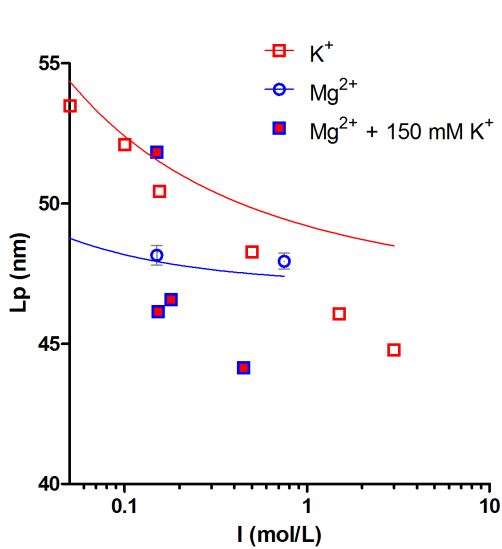

FIG. 4: Influence of the addition of divalent ions such as Mg$^{2+}$ on dsDNA persistence length.

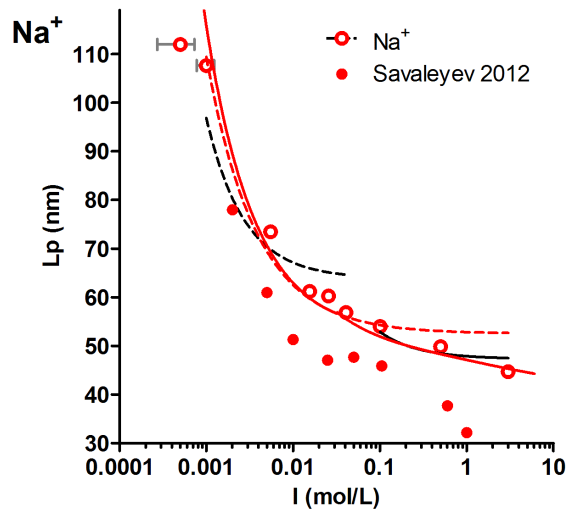

FIG. 5: Influence of Na$^+$ on dsDNA persistence length: comparison of our experimental data (empty circles) with simulation results obtained by Savelyev [9] (filled circle) with fitting curves corresponding to OSF (black line), OSFM (black dotted line), NO (red dotted line) and TS (red line) formulas.

|  | Ions | I range | Fixed values | | | Free parameters (nm) | |
|---|---|---|---|---|---|---|---|
|  |  |  | C | δ | β=2(1-δ) | $L_p^0$ | $R_{dsDNA}$ |
| OSF | Mg$^{2+}$ | > 0.1 M | 0.559 | 1 | - | 45.7 | - |
|  | Ca$^{2+}$ | > 0.1 M | 0.559 | 1 | - | 45.4 | - |
|  | Put$^{2+}$ | > 0.1 M | 0.559 | 1 | - | 46.9 | - |
| OSFM | Mg$^{2+}$ | [0.0009 M ;0.1 M] | 0.008 | 1 | - | 53.6 | - |
|  | Ca$^{2+}$ | [0.0009 M ;0.1 M] | 0.008 | 1 | - | 53.8 | - |
|  | Put$^{2+}$ | [0.0009 M ;0.1 M] | 0.008 | 1 | - | 57.4 | - |
| NO | Mg$^{2+}$ | > 0.0009 M | 0.238 | 0.636 | 0.728 | 47.1 | 1.05 |
|  | Ca$^{2+}$ | > 0.0009 M | 0.238 | 0.636 | 0.728 | 46.0 | 1.20 |
|  | Put$^{2+}$ | > 0.0009 M | 0.238 | 0.636 | 0.728 | 46.2 | 1.57 |

TABLE II: Analysis of divalent ions results using the fitting function of the OSF and OSFM theories, $L_P = L_P^0 + C/I^\delta$ with $L_P^0$ as the only free parameter, and the NO fitting function, $L_P = L_P^0 + C/I^\delta R_{\rm DNA}^\beta$ with $L_P^0$ and $R_{\rm DNA}$ as free parameters.

|  | Ions | I range | Fixed values | | | Free parameters (nm) | |
|---|---|---|---|---|---|---|---|
|  |  |  | C | δ | β=2(1-δ) | $L_p^0$ | $R_{dsDNA}$ |
| OSF | Li$^+$ | > 0.1 M | 0.559 | 1 | - | 44.1 |  |
|  | Na$^+$ | > 0.1 M | 0.559 | 1 | - | 47.3 |  |
|  | K$^+$ | > 0.1 M | 0.559 | 1 | - | 46.2 |  |
|  | TMA$^+$ PEG | > 0.1 M | 0.559 | 1 | - | 56.4 |  |
|  | TEA$^+$ PEG | > 0.1 M | 0.559 | 1 | - | 52.5 |  |
| OSFM | Li$^+$ | [0.0009 M ;0.1 M] | 0.033 | 1 | - | 60.0 |  |
|  | Na$^+$ | [0.0009 M ;0.1 M] | 0.033 | 1 | - | 63.8 |  |
|  | K$^+$ | [0.0009 M ;0.1 M] | 0.033 | 1 | - | 66.4 |  |
|  | TMA$^+$ PEG | [0.0009 M ;0.1 M] | 0.033 | 1 | - | 73.2 |  |
|  | TEA$^+$ PEG | [0.0009 M ;0.1 M] | 0.033 | 1 | - | 65.1 |  |
| NO | Li$^+$ | > 0.0009 M | 0.404 | 0.755 | 0.490 | 50.8 | 0.48 |
|  | Na$^+$ | > 0.0009 M | 0.404 | 0.755 | 0.490 | 52.5 | 0.58 |
|  | K$^+$ | > 0.0009 M | 0.404 | 0.755 | 0.490 | 51.1 | 0.76 |
|  | TMA$^+$ PEG | > 0.0009 M | 0.404 | 0.755 | 0.490 | 61.1 | 0.50 |
|  | TEA$^+$ PEG | > 0.0009 M | 0.404 | 0.755 | 0.490 | 57.7 | 0.36 |
| Trizac | Li$^+$ | > 0.0009 M |  |  |  | 41 | 1 |
|  | Na$^+$ | > 0.0009 M |  |  |  | 41 | 0.85 |
|  | K$^+$ | > 0.0009 M |  |  |  | 41 | 0.8 |
|  | TMA$^+$ PEG | > 0.0009 M |  |  |  | 51 | 0.95 |
|  | TEA$^+$ PEG | > 0.0009 M |  |  |  | 47 | 1.1 |

TABLE III: Analysis of monovalent ion results using the fitting function of the OSF and OSFM theories, $L_P = L_P^0 + C/I^\delta$ with $L_P^0$ as the only free parameter, the NO fitting function $L_P = L_P^0 + C/I^\delta R_{\rm DNA}^\beta$ with $L_P^0$ and $R_{\rm DNA}$ as free parameters, and the TS fitting function (see main text).

| [Li+] (mM) | I(M) | particleNumber | meanSample | sdr.errorBOOT | Lp(nm) | SD(lp) |
|---|---|---|---|---|---|---|
| 0 | 0,0005 | 2179 | 247,72 | 0,25 | 102,10 | 0,52 |
| 0,5 | 0,001 | 2629 | 246,40 | 0,27 | 99,35 | 0,56 |
| 1 | 0,0015 | 2341 | 241,78 | 0,25 | 90,22 | 0,47 |
| 2,5 | 0,003 | 2051 | 233,76 | 0,21 | 76,17 | 0,34 |
| 5 | 0,0055 | 1759 | 229,44 | 0,29 | 69,57 | 0,42 |
| 10 | 0,0105 | 1535 | 222,91 | 0,26 | 60,83 | 0,31 |
| 15 | 0,0155 | 1814 | 221,18 | 0,25 | 58,77 | 0,29 |
| 20 | 0,0205 | 1705 | 221,17 | 0,27 | 58,75 | 0,32 |
| 25 | 0,0255 | 1692 | 219,61 | 0,24 | 56,99 | 0,26 |
| 30 | 0,0305 | 1672 | 217,80 | 0,24 | 55,05 | 0,25 |
| 40 | 0,0405 | 1418 | 217,65 | 0,28 | 54,89 | 0,29 |
| 50 | 0,0505 | 1397 | 214,94 | 0,25 | 52,23 | 0,24 |
| 100 | 0,1005 | 690 | 211,06 | 0,44 | 48,86 | 0,35 |
| 155 | 0,1555 | 775 | 212,15 | 0,45 | 49,75 | 0,37 |
| 500 | 0,5005 | 508 | 210,53 | 0,55 | 48,43 | 0,43 |
| 1500 | 1,5005 | 616 | 204,33 | 0,61 | 44,28 | 0,34 |
| 3000 | 3,0005 | 612 | 202,79 | 0,55 | 43,46 | 0,28 |
| 6000 | 6,0005 | 613 | 195,48 | 0,48 | 40,72 | 0,12 |

| [Na+] (mM) | I(M) | particleNumber | meanSample | sdr.errorBOOT | lp(nm) | SD(lp) |
|---|---|---|---|---|---|---|
| 0 | 0,0005 | 890 | 252,23 | 0,31 | 111,95 | 0,70 |
| 0,5 | 0,001 | 880 | 250,29 | 0,29 | 107,63 | 0,64 |
| 5 | 0,0055 | 851 | 232,05 | 0,28 | 73,48 | 0,43 |
| 15 | 0,0155 | 790 | 223,21 | 0,29 | 61,20 | 0,36 |
| 25 | 0,0255 | 779 | 222,43 | 0,28 | 60,25 | 0,34 |
| 40 | 0,0405 | 649 | 219,52 | 0,30 | 56,90 | 0,33 |
| 100 | 0,1005 | 600 | 216,78 | 0,32 | 54,01 | 0,33 |
| 500 | 0,5005 | 660 | 212,26 | 0,30 | 49,84 | 0,25 |
| 3000 | 3,0005 | 573 | 205,11 | 0,32 | 44,73 | 0,19 |

| [K+] (mM) | I(M) | particleNumber | meanSample | sdr.errorBOOT | lp(nm) | SD(lp) |
|---|---|---|---|---|---|---|
| 0 | 0,0005 | 2568 | 251,98 | 0,27 | 111,39 | 0,61 |
| 0,5 | 0,001 | 1652 | 251,19 | 0,25 | 109,61 | 0,55 |
| 1 | 0,0015 | 3096 | 246,53 | 0,16 | 99,61 | 0,34 |
| 2,5 | 0,003 | 1756 | 240,07 | 0,20 | 87,04 | 0,37 |
| 5 | 0,0055 | 2416 | 232,91 | 0,17 | 74,82 | 0,27 |
| 10 | 0,0105 | 1124 | 226,26 | 0,28 | 65,12 | 0,38 |
| 15 | 0,0155 | 1140 | 221,79 | 0,30 | 59,48 | 0,36 |
| 20 | 0,0205 | 618 | 219,42 | 0,31 | 56,79 | 0,34 |
| 25 | 0,0255 | 1172 | 217,62 | 0,26 | 54,87 | 0,27 |
| 30 | 0,0305 | 1196 | 219,71 | 0,26 | 57,10 | 0,29 |
| 40 | 0,0405 | 792 | 218,63 | 0,45 | 55,93 | 0,48 |
| 50 | 0,0505 | 1132 | 216,25 | 0,30 | 53,48 | 0,30 |
| 100 | 0,1005 | 1134 | 214,81 | 0,24 | 52,10 | 0,23 |
| 155 | 0,1555 | 1086 | 212,95 | 0,25 | 50,43 | 0,22 |
| 500 | 0,5005 | 1185 | 210,31 | 0,28 | 48,27 | 0,22 |
| 1500 | 1,5005 | 1022 | 207,24 | 0,26 | 46,06 | 0,17 |
| 3000 | 3,0005 | 956 | 205,21 | 0,28 | 44,79 | 0,17 |

| [TMA+] (mM) | I(M) | particleNumber | meanSample | sdr.errorBOOT | lp(nm) | SD(lp) |
|---|---|---|---|---|---|---|
| 0 | 0,0005 | 1774 | 253,03 | 0,28 | 113,78 | 0,63 |
| 0,5 | 0,001 | 1044 | 248,49 | 0,33 | 103,71 | 0,70 |
| 1 | 0,0015 | 1882 | 245,30 | 0,26 | 97,11 | 0,52 |
| 2,5 | 0,003 | 912 | 239,39 | 0,22 | 85,80 | 0,40 |
| 5 | 0,0055 | 1788 | 235,62 | 0,22 | 79,24 | 0,37 |
| 10 | 0,0105 | 1176 | 230,14 | 0,27 | 70,60 | 0,39 |
| 15 | 0,0155 | 1002 | 226,55 | 0,29 | 65,52 | 0,39 |
| 20 | 0,0205 | 1030 | 224,82 | 0,30 | 63,23 | 0,39 |
| 25 | 0,0255 | 1089 | 225,76 | 0,31 | 64,46 | 0,42 |
| 30 | 0,0305 | 810 | 222,25 | 0,33 | 60,03 | 0,39 |
| 40 | 0,0405 | 941 | 224,17 | 0,38 | 62,40 | 0,48 |
| 50 | 0,0505 | 1055 | 223,28 | 0,33 | 61,29 | 0,40 |
| 100 | 0,1005 | 1101 | 223,36 | 0,40 | 61,39 | 0,49 |
| 155 | 0,1555 | 1606 | 224,47 | 0,24 | 62,78 | 0,31 |
| 300 | 0,3005 | 823 | 227,24 | 0,27 | 66,45 | 0,37 |
| 500 | 0,5005 | 1653 | 224,13 | 0,26 | 62,35 | 0,33 |
| 700 | 0,7005 | 982 | 223,89 | 0,22 | 62,05 | 0,27 |
| 1500 | 1,5005 | 1592 | 216,17 | 0,27 | 53,41 | 0,26 |
| 3000 | 3,0005 | 972 | 209,25 | 0,34 | 47,46 | 0,25 |

| [TMA+] PEG (mM) | I(M) | particleNumber | meanSample | sdr.errorBOOT | lp(nm) | SD(lp) |
|---|---|---|---|---|---|---|
| 0 | 0,0005 | 1117 | 253,70 | 0,32 | 115,32 | 0,75 |
| 0,5 | 0,001 | 1324 | 250,70 | 0,26 | 108,53 | 0,58 |
| 1 | 0,0015 | 1134 | 248,73 | 0,25 | 104,24 | 0,53 |
| 2,5 | 0,003 | 1118 | 240,96 | 0,25 | 88,68 | 0,47 |
| 5 | 0,0055 | 917 | 235,82 | 0,32 | 79,56 | 0,54 |
| 10 | 0,0105 | 866 | 231,87 | 0,29 | 73,21 | 0,45 |
| 15 | 0,0155 | 583 | 230,51 | 0,30 | 71,14 | 0,44 |
| 25 | 0,0255 | 856 | 228,56 | 0,30 | 68,30 | 0,42 |
| 40 | 0,0405 | 832 | 225,44 | 0,32 | 64,04 | 0,41 |
| 100 | 0,1005 | 793 | 224,74 | 0,34 | 63,13 | 0,44 |
| 300 | 0,3005 | 893 | 222,65 | 0,26 | 60,51 | 0,32 |
| 500 | 0,5005 | 901 | 220,95 | 0,32 | 58,50 | 0,37 |
| 700 | 0,7005 | 572 | 218,74 | 0,34 | 56,05 | 0,37 |
| 1500 | 1,5005 | 573 | 213,99 | 0,36 | 51,36 | 0,32 |

| [TEA+] (mM) | I(M) | particleNumber | meanSample | sdr.errorBOOT | lp(nm) | SD(lp) |
|---|---|---|---|---|---|---|
| 0 | 0,0005 | 3488 | 251,91 | 0,20 | 111,23 | 0,44 |
| 0,5 | 0,001 | 1689 | 253,82 | 0,24 | 115,60 | 0,56 |
| 1 | 0,0015 | 3905 | 249,47 | 0,17 | 105,83 | 0,37 |
| 2,5 | 0,003 | 2916 | 239,71 | 0,17 | 86,38 | 0,30 |
| 5 | 0,0055 | 3583 | 240,61 | 0,16 | 88,02 | 0,30 |
| 10 | 0,0105 | 1274 | 239,37 | 0,22 | 85,75 | 0,39 |
| 15 | 0,0155 | 1906 | 234,18 | 0,17 | 76,86 | 0,28 |
| 20 | 0,0205 | 1774 | 232,21 | 0,16 | 73,73 | 0,24 |
| 25 | 0,0255 | 1861 | 233,33 | 0,23 | 75,49 | 0,37 |
| 30 | 0,0305 | 1511 | 227,60 | 0,20 | 66,95 | 0,27 |
| 40 | 0,0405 | 1836 | 224,95 | 0,24 | 63,40 | 0,31 |
| 50 | 0,0505 | 1729 | 224,54 | 0,26 | 62,87 | 0,34 |
| 100 | 0,1005 | 1359 | 224,03 | 0,19 | 62,22 | 0,23 |
| 155 | 0,1555 | 1567 | 218,10 | 0,20 | 55,37 | 0,21 |
| 200 | 0,2005 | 1291 | 216,79 | 0,27 | 54,02 | 0,27 |
| 300 | 0,3005 | 1201 | 211,86 | 0,33 | 49,51 | 0,27 |

| [TEA+] PEG (mM) | I(M) | particleNumber | meanSample | sdr.errorBOOT | lp(nm) | SD(lp) |
|---|---|---|---|---|---|---|
| 0 | 0,0005 | 2029 | 249,43 | 0,24 | 105,75 | 0,53 |
| 0,5 | 0,001 | 2713 | 247,94 | 0,21 | 102,56 | 0,45 |
| 1 | 0,0015 | 419 | 241,17 | 0,51 | 89,07 | 0,96 |
| 2,5 | 0,003 | 588 | 233,96 | 0,36 | 76,50 | 0,57 |
| 5 | 0,0055 | 2234 | 231,95 | 0,24 | 73,32 | 0,37 |
| 10 | 0,0105 | 579 | 225,65 | 0,45 | 64,31 | 0,60 |
| 15 | 0,0155 | 514 | 223,16 | 0,48 | 61,14 | 0,60 |
| 20 | 0,0205 | 2031 | 225,12 | 0,23 | 63,62 | 0,30 |
| 40 | 0,0405 | 1882 | 224,30 | 0,27 | 62,56 | 0,34 |
| 155 | 0,1555 | 1591 | 220,14 | 0,31 | 57,58 | 0,34 |
| 300 | 0,3005 | 1327 | 215,21 | 0,34 | 52,48 | 0,32 |

TABLE IV: Results obtained with the monovalent ions.

| [Mg2+] (mM) | I(M) | particleNumber | meanSample | sdr.errorBOOT | lp(nm) | SD(lp) |
|---|---|---|---|---|---|---|
| 0 | 0,0005 | 541 | 251,97 | 0,51 | 111,36 | 1,16 |
| 0,25 | 0,00125 | 592 | 226,21 | 0,39 | 65,06 | 0,52 |
| 0,75 | 0,00275 | 480 | 218,84 | 0,39 | 56,16 | 0,42 |
| 2 | 0,0065 | 483 | 215,95 | 0,33 | 53,19 | 0,32 |
| 10 | 0,0305 | 447 | 211,64 | 0,34 | 49,32 | 0,28 |
| 50 | 0,1505 | 433 | 210,17 | 0,45 | 48,16 | 0,35 |
| 250 | 0,7505 | 362 | 209,89 | 0,38 | 47,95 | 0,29 |
| [Ca2+] (mM) | I(M) | particleNumber | meanSample | sdr.errorBOOT | lp(nm) | SD(lp) |
| 0 | 0,0005 | 4878 | 249,37 | 0,19 | 105,62 | 0,41 |
| 0,075 | 0,000725 | 1825 | 236,85 | 0,25 | 81,32 | 0,43 |
| 0,15 | 0,00095 | 2541 | 230,84 | 0,22 | 71,64 | 0,33 |
| 0,25 | 0,00125 | 1792 | 223,44 | 0,25 | 61,49 | 0,31 |
| 0,5 | 0,002 | 1631 | 221,92 | 0,29 | 59,64 | 0,35 |
| 0,75 | 0,00275 | 1588 | 219,41 | 0,28 | 56,77 | 0,30 |
| 1 | 0,0035 | 1339 | 217,97 | 0,33 | 55,23 | 0,34 |
| 2 | 0,0065 | 1525 | 214,22 | 0,26 | 51,56 | 0,24 |
| 5 | 0,0155 | 1505 | 211,47 | 0,28 | 49,18 | 0,22 |
| 10 | 0,0305 | 1531 | 210,68 | 0,33 | 48,56 | 0,26 |
| 20 | 0,0605 | 1453 | 211,57 | 0,25 | 49,27 | 0,21 |
| 50 | 0,1505 | 1414 | 208,44 | 0,26 | 46,89 | 0,19 |
| 100 | 0,3005 | 1493 | 207,70 | 0,32 | 46,37 | 0,22 |
| 250 | 0,7505 | 1474 | 207,62 | 0,28 | 46,32 | 0,19 |
| 500 | 1,5005 | 2247 | 208,50 | 0,23 | 46,93 | 0,17 |
| 1000 | 3,0005 | 1295 | 208,65 | 0,29 | 47,03 | 0,21 |
| [Put2+] (mM) | I(M) | particleNumber | meanSample | sdr.errorBOOT | lp(nm) | SD(lp) |
| 0 | 0,0005 | 2551 | 248,10 | 0,26 | 102,90 | 0,55 |
| 0,075 | 0,000725 | 1940 | 244,76 | 0,26 | 96,02 | 0,53 |
| 0,15 | 0,00095 | 3104 | 234,63 | 0,20 | 77,59 | 0,33 |
| 0,25 | 0,00125 | 1969 | 227,57 | 0,17 | 66,91 | 0,24 |
| 0,5 | 0,002 | 1712 | 222,26 | 0,20 | 60,04 | 0,24 |
| 0,75 | 0,00275 | 1688 | 219,76 | 0,20 | 57,16 | 0,22 |
| 1 | 0,0035 | 1675 | 218,62 | 0,20 | 55,91 | 0,22 |
| 2 | 0,0065 | 1527 | 216,19 | 0,21 | 53,43 | 0,21 |
| 5 | 0,0155 | 979 | 213,91 | 0,24 | 51,28 | 0,22 |
| 10 | 0,0305 | 939 | 212,80 | 0,21 | 50,30 | 0,18 |
| 20 | 0,0605 | 912 | 212,20 | 0,23 | 49,79 | 0,19 |
| 50 | 0,1505 | 901 | 211,08 | 0,30 | 48,87 | 0,24 |
| 100 | 0,3005 | 672 | 212,20 | 0,34 | 49,79 | 0,29 |
| 250 | 0,7505 | 1074 | 210,28 | 0,33 | 48,24 | 0,25 |
| 500 | 1,5005 | 1032 | 209,22 | 0,25 | 47,45 | 0,18 |

TABLE V: Results obtained with the divalent ions.


\end{document}